\documentclass[11pt,twoside]{article} 
\usepackage{asp2004}
\usepackage{epsf}
\usepackage{psfig}
\usepackage{lscape} 

\begin{document} 
\title{A NICMOS Direct Imaging Search for Giant Planets around 
the Single White Dwarfs in the Hyades}
\author{S.~Friedrich,$^{1}$ H.~Zinnecker,$^{2}$ W. Brandner,$^{3}$
    S. Correia,$^{4}$ and M. McCaughrean$^{2}$} 
\affil{$^1$Max-Planck-Institut f\"ur Extraterrestrische Physik, 
    Giessenbachstr., 85748 Garching, Germany\\
    $^2$Astrophysikalisches Institut Potsdam, An der 
    Sternwarte 16, 14482 Potsdam, Germany\\
    $^3$Max-Planck-Institut f\"ur Astronomie, Auf dem
    K\"onigstuhl, 69117 Heidelberg, Germany\\ 
    $^4$Observatoire de Paris, LESIA, UMR 8109, 92190 Meudon, France\\}
\begin{abstract} 
We report preliminary results from our search for massive giant planets (6-12
Jupiter masses) around
the known seven single white dwarfs in the Hyades cluster at sub-arcsec
separations. At an age of 625 Myr, the white dwarfs had progenitor masses of
about 3 solar masses, and massive gaseous giant planets should have formed in
the massive circumstellar disks around these ex-Herbig A0 stars, probably at
orbital separations similar or slightly larger than that of Jupiter. Such
planets would have survived the post-Main-Sequence mass loss of the parent
star and would have migrated outward adiabatically to a distance of about 25
AU. At the distance of the Hyades (45 pc) this corresponds to an angular
separation of 0\farcs5. J and H magnitudes of these giants are in the 
range of 20.5-23.3 mag, which can be resolved with NICMOS.

The achieved sensitivities and contrast ratios agree well with
simulations. Preliminary evaluation of the NICMOS data set did not reveal any
evidence for neither planetary mass companions with masses down to 
about 10 Jupiter masses nor brown dwarfs around any of the seven white dwarfs 
for separations larger than 0\farcs5.

\end{abstract}

\section{Planets around White Dwarfs}
Since the discovery of the first extrasolar planet by Mayor \& Queloz
(1995) the
number of known extrasolar planets has grown to more than 100. Most
discoveries are based on measurements of radial velocities of stellar lines.
Due to the great contrast in optical brightness
between the main sequence star and its planet(s) no planet
has so far been seen directly.
A much better spectral contrast between an extrasolar giant planet 
and a star can be achieved in the IR band, where the planet's thermal
emission peaks. The contrast can be increased 
further if one searches
for planets around white dwarfs, which are about a factor of ten smaller
than a giant planet ($F\sim R^2$) and typically $10^3$ to $10^4$ times less
luminous than main sequence stars of spectral type G or K. 

For our search we have chosen the only young cluster in the near vicinity 
with a large enough sample of white dwarfs: the Hyades. 
Compared to isolated white dwarfs this has the 
advantage that the age of the white dwarfs is known from the turn-off age 
of the Hyades (625 Myr, Perryman et al. 1998). This also implies that 
any planet will also be of the same age. A Hyades main sequence star 
with 3 M$_\odot$ (the turn-off mass) will eventually become a 
0.66 M$_\odot$ white dwarf (Weidemann 2000). All
single stars with masses above the turn-off mass and below about 8 M$_\odot$ 
have already left the main sequence and will develop into or have already
become white dwarfs. While planets closer than 3-5 AU will probably not 
survive this post-main-sequence evolution (and will migrate
inwards and merge) planets farther away will survive (Duncan \& Lissauer 
1998; Burleigh et al. 2002), and their semimajor axis will
increase by a factor $M_{\rm initial}$/$M_{\rm final}$ 
as the central star loses mass. Thus any giant planet in a
Jupiter-like orbit ($\ge$ 5 AU) will migrate outwards to a new equilibrium
radius of at least:\\ 
\centerline{R=5 AU$\times$(3M$_\odot$ / 0.66M$_\odot$) $\approx$ 23 AU}
At a distance of the Hyades (45 pc) 23 AU correspond to 0\farcs5. 
Furthermore, a giant planet with mass somewhat below 13 Jupiter 
masses (which is one of the
definition for ``planet'') orbiting a white dwarf in the Hyades has
{\bf a
near-infrared magnitude difference of only about 7 mag}. Both is feasible 
with NICMOS in the F110W and F160W filters with spatial
resolutions of 0\farcs1 and 0\farcs15, respectively.  

\section{Observations and Data Reduction}
The seven single white dwarfs (HZ4, LB227, VR7, VR16, HZ7, HZ14, LP475-242) 
have been observed with HST/NICMOS at 
two roll angles separated by 20 degrees through the F110W and F160W filters. 
Two and four dithered frames, respectively, of 320s integration time each
were recorded for each roll angle and each target in these filters, which
corresponds to a total of 
one and two orbits of HST time per target, respectively. 
The single dithered frames were reduced
using the HST pipeline, then oversampled
by a factor two using cubic spline before registration and a FFT-shift based
combination for each roll angle.
Subsequent bad-pixel masking was further applied in order to compensate for the
non-optimal bad-pixel mask
available through the pipeline. For that purpose, remaining not 
corrected bad-pixels were identified using a
criterion based on the deviation from the median of all frames available 
for each target.
In each filter, the resulting frames corresponding to different roll angles 
were subtracted from each other.
This step was done after the use of a 2D cross-correlation for the 
registration of these frames, which provided
a sub-pixel accuracy, and a FFT-shift based combination. The result 
obtained for HZ7 is presented in Fig. 1 as an example.

\begin{figure}[!ht]
\plotfiddle{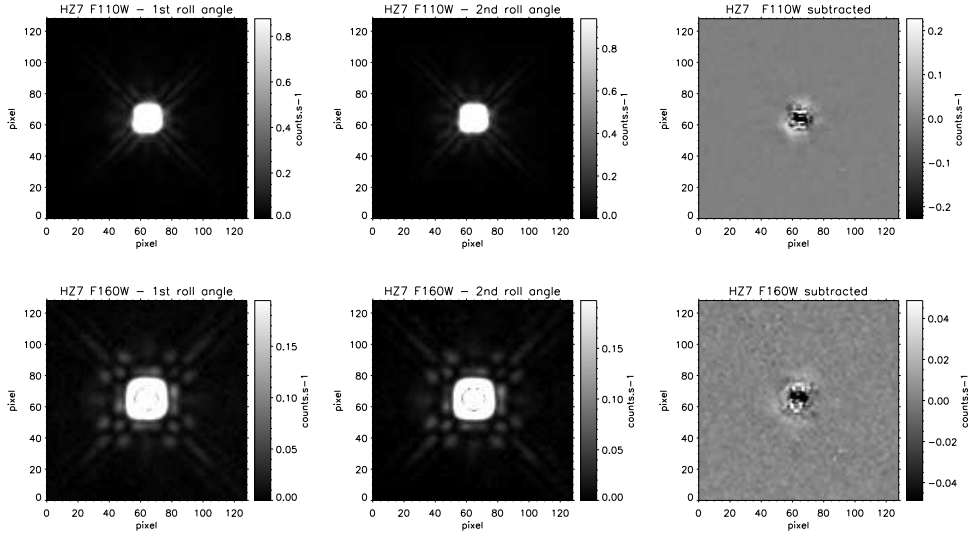}{6.5cm}{90}{50}{50}{190}{-10}
\caption{The observations of HZ7 at two different roll angles 
and the result of their subtractions in the F110W filter (upper panel) and
the F160W filter (lower panel). The field of view is about 
$2\farcs75\times 2\farcs75$ (0\farcs0215 pixel size). 
Maximum cuts are at 2\%
of the peak intensity for the observations, and $\pm$0.5\% for the subtracted 
frames}
\end{figure}
\begin{figure}[!ht]
\plotfiddle{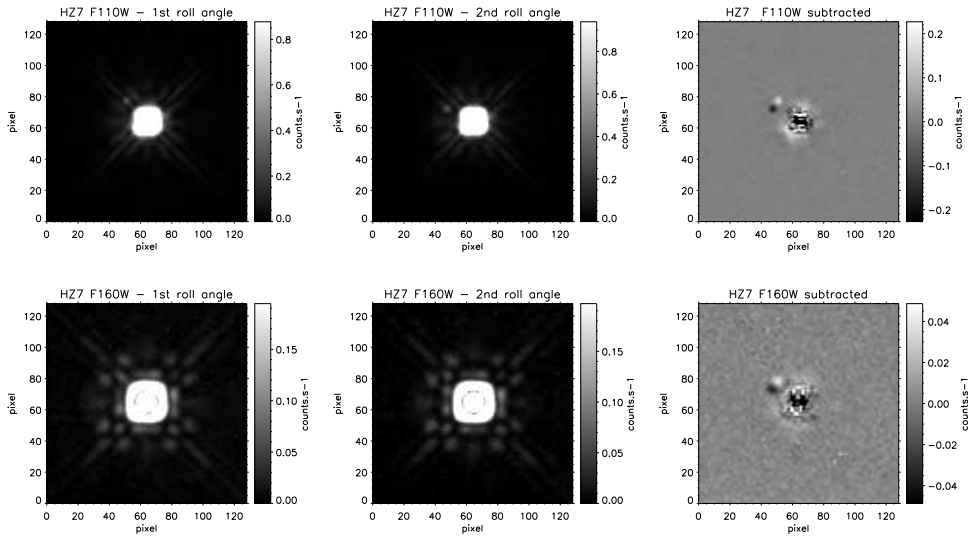}{6.5cm}{90}{50}{50}{190}{-10}
\caption{Same as Fig. 1 with an artificial planetary mass companion
added (6 mag brightness difference, 0\farcs4 separation)}
\end{figure}

\section{Method Validation and Contrast Ratio}
We validated the method used in the data reduction through simulations. 
For this purpose a fake companion of given brightness was introduced 
in each frame as a scaled version of the target, adding the appropriate noise,
and checking that we could detect the signature of the companion 
after all reduction steps described above.
Fig.~2 shows the results of such a simulation for both filters and the 
same target as in Fig.~1, with a planet at 0\farcs4
separation and 6 magnitudes difference in brightness. One can 
easily notice the pair of negative and positive images of the
companion at the given position from the star in the subtracted frame.
Such a simulation 
allows to estimate the limiting difference in brightness 
as a function of separation and given SNR of the detection. 
Fig.~3 shows the result for the same target as in Fig.~1 in each filter 
and a detection
threshold of 5$\sigma$. The uncertainties are estimated from the scatter 
obtained using four different position angles of the companion.
According to the COND models of Baraffe et al. (2003), 
the limiting mass of the planetary mass companion
that we are able to detect is typically of about 10 Jupiter masses 
at 0\farcs5 separation.

\begin{figure}[!ht]
\plotfiddle{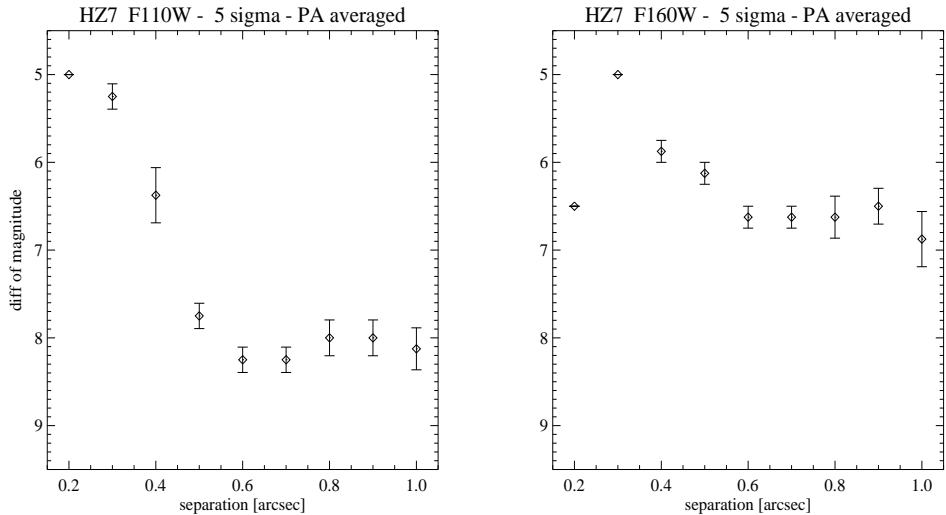}{6.5cm}{90}{48}{50}{200}{-55}
\caption{Estimated sensitivity limits as a function of separation for HZ7.
Uncertainties are estimated from the scatter obtained
using four different position angles of the companion.}
\end{figure}

\section{Conclusion}
We did not find any evidence for neither planetary mass companions with masses
down to about 10 Jupiter masses nor brown dwarfs around any of the seven white
dwarfs in the Hyades for separations larger than 0\farcs5. This could mean
that extrasolar giant planets are fainter than expected, that accompanying 
planets are too close to their parent star to be resolved, that massive 
Jupiter-like planets in wide orbits around 3-4 M$_{\odot}$ main sequence 
stars are not common or that these planets do not remain bound to white dwarfs,
for example due to planet-planet perturbations during mass loss
(Debes \& Sigurdsson 2002).


\end{document}